\documentclass[sigconf]{acmart}
\setcopyright{none} % to remove the copyright notice
\renewcommand\footnotetextcopyrightpermission[1]{}

\usepackage{graphicx}
\usepackage{subfigure}
\usepackage{multirow}
\usepackage{multicol}
\usepackage[breakable]{tcolorbox}
\usepackage{pifont}
\usepackage{enumitem}

\usepackage[normalem]{ulem}
\useunder{\uline}{\ul}{}
\usepackage{threeparttable}

\usepackage{xspace}
\usepackage{graphicx}
\usepackage{textcomp}
\usepackage{xcolor}
\usepackage{hyperref}
\usepackage{graphicx}
\usepackage{subfigure}
\usepackage{multirow}
\usepackage{multicol}
\usepackage[breakable]{tcolorbox}
\usepackage{pifont}
\usepackage{enumitem}
\usepackage{amsmath}

\usepackage{bbm}

\usepackage{makecell}
\usepackage{tcolorbox}
\usepackage{booktabs}

\usepackage{listings}

\usepackage{algorithm}
\usepackage{algpseudocode}

\usepackage{tikz}
\newcommand*{\circled}[1]{\lower.7ex\hbox{\tikz\draw (0pt, 0pt)%
    circle (.5em) node {\makebox[1em][c]{\small #1}};}}

\renewcommand{\paragraph}[1]{\vskip 0.01in \noindent {\bf #1.}}

\makeatletter
\DeclareRobustCommand\onedot{\futurelet\@let@token\@onedot}
\def\@onedot{\ifx\@let@token.\else.\null\fi\xspace}
 
\def\eg{\emph{e.g}\onedot} \def\Eg{\emph{E.g}\onedot}
\def\ie{\emph{i.e}\onedot} 
 
\def\etc{\emph{etc}\onedot}

\DeclareRobustCommand{\method}{SEAlign\xspace}
\makeatother

\AtBeginDocument{%
  }

\setcopyright{acmlicensed}
\copyrightyear{2018}
\acmYear{2018}
\acmDOI{XXXXXXX.XXXXXXX}
\acmConference[Conference acronym 'XX]{Make sure to enter the correct
  conference title from your rights confirmation email}{June 03--05,
  2018}{Woodstock, NY}

\acmISBN{978-1-4503-XXXX-X/2018/06}

\begin{document}

\title{\method: Alignment Training for Software Engineering Agent}

\author{Kechi Zhang\textsuperscript{$\dagger$}, Huangzhao Zhang, Ge Li\textsuperscript{$\dagger$} \footnotemark[1], Jinliang You\textsuperscript{$\dagger$}, Jia Li\textsuperscript{$\dagger$}, Yunfei Zhao\textsuperscript{$\dagger$}, Zhi Jin\textsuperscript{$\dagger$} \footnotemark[1]}
\affiliation{%
\institution{\textsuperscript{$\dagger$}Key Lab of High Confidence Software Technology (PKU), Ministry of Education, \\
School of Computer Science, Peking University\\}
  \country{China}
  }
\email{{zhangkechi,lige,zhijin}@pku.edu.cn}

\renewcommand{\shortauthors}{Trovato et al.}

\begin{abstract}
\renewcommand{\thefootnote}{\fnsymbol{footnote}}
\footnotetext[1]{Corresponding authors.}
\renewcommand{\thefootnote}{\arabic{footnote}}

Recent advances in code generation models have demonstrated impressive capabilities in automating software development tasks, yet these models still struggle in real-world software engineering scenarios. 
Although current training methods, particularly post-training, excel at solving competitive programming problems, they fail to adequately prepare models for the complexities of practical software development. 
This misalignment raises the critical question: Are existing alignment training methods well suited for real-world software engineering tasks?
In this study, we identify this issue and propose \method, a novel alignment framework designed to bridge the gap between code generation models and real-world software development tasks. \method leverages the unique characteristics of software engineering processes, including high-quality workflow steps, to enhance model capabilities.
Our framework further employs Monte Carlo Tree Search for fine-grained alignment in multi-step decision processes, followed by preference optimization on critical actions to ensure models meet real-world requirements.
We evaluate \method on three standard agentic benchmarks for real-world software engineering, including HumanEvalFix, SWE-Bench-Lite, and SWE-Bench-Verified. 
Experimental results demonstrate state-of-the-art performance with minimal training overhead. 
In addition, we develop an agent-based software development platform using \method, which successfully automates the creation of several small applications. 
Human evaluations of these applications highlight significant improvements in both task performance and user experience.
Our findings underscore the potential of \method to accelerate the adoption of large code models in real-world software development.
We believe that this research makes a meaningful step towards fully automated software engineering. 
\footnote{Code and additional details are available at \cite{casestudy}.}
%  \url{https://anonymous.4open.science/r/SWEAlign/}

\end{abstract}

\received{20 February 2007}
\received[revised]{12 March 2009}
\received[accepted]{5 June 2009}
\maketitle

\section{Introduction}
Code generation has become a pivotal area in artificial intelligence, with models automating essential software development tasks and significantly boosting developer productivity. Recent advances in code models (such as GPT-4 \cite{GPT-4}, LLaMA \cite{Llama}, and DeepSeekCoder \cite{guo2024deepseek} ), have shown impressive abilities to generate functional and effective code. These models have greatly improved automation in various code-related tasks, making software development more efficient.

Current code generation models, \ie, Large Language Models (LLMs), typically follow a two-stage training process: \ding{182} pre-training, where the model learns basics of programming languages and foundational skills, and \ding{183} post-training, where the model learns to follow specific instructions and is aligned with human behaviors and preferences.
Recent alignment training techniques for code models, such as CodeDPO \cite{codedpo} and LIMO \cite{ye2025limo}, have been proven highly effective, achieving significant improvements with minimal data and computational resources in certain scenarios.
However, the post-training phase of existing code generation models focuses mainly on solving competitive programming contest problems from platforms like LeetCode \cite{leetcode} and Codeforces \cite{codeforces} -- there is a significant gap between algorithmic problems and real-world software development scenarios.
Practical software engineering demands a composite of various skills, including the retrieval of information from complex repositories and the flexible application of existing algorithms to current code contexts, \etc, but existing post-training solutions often perform poorly.
Bridging this gap is crucial for accelerating the adoption of code models in software development automation.

Despite the rapid advancements in model capabilities which have surpassed the majority of human programmers on competitive coding benchmarks, these models often struggle in real-world software development scenarios as aforementioned. Benchmarks such as SWE-bench \cite{swebench}, which simulates real-world scenarios of issue fixing, reveal that many open-source code models perform less satisfactorily.
This discrepancy raises an important question: Are current code model training schemes (specifically post-training techniques) capable to aligned the models with the skills required for real-world software development well?
To answer the question in a brief and straightforward manner, we conduct a preliminary investigation, evaluating Qwen2.5-Coder-Instruct-14B \cite{qwencoder} (which is a powerful code model extensively post-trained) and Openhands \cite{wang2024openhands} (which is a popular agentic framework) against SWE-Bench-Lite \cite{swebench} (which is a benchmark simulating real-world software engineering tasks).
The experiment shows that the model could only solve 3.7\% of problems within the benchmark.
Diving into the failure cases, we find that the incapability of the model is mainly caused by \ding{182} the poor instruction following ability, \ding{183} the incorrect tool selection and usage, and \ding{184} infinite agentic loops (refer to Section \ref{sec:failuremisalign} for more analysis).
Our preliminary investigation results directly reveal that existing post-training methods do not adequately prepare models for the complexities of software engineering tasks, especially within agentic framework.

To address this issue, we propose \textbf{\method}, a novel framework designed to align code models with real-world software development agentic workflows. 
\method focuses on the unique characteristics of software engineering, enhancing models’ ability to handle complex tasks. 
In general, \method enables code models to follow instructions and use developing tools correctly by tuning them under real-world tasks.
Specifically, we collect high-quality samples of software development agentic trajectories (\ie, decision processes), identify critical action steps in the workflow, and force the model to produce ``good'' behaviors by alignment.
\method uses Monte Carlo Tree Search (MCTS) to provide detailed scoring and alignment for these multi-step decision processes, followed by preference optimization to ensure the model meets real-world requirements.

We evaluate \method on three standard benchmarks for real-world software engineering tasks: HumanEvalFix \cite{HumanEvalFix}, SWE-Bench-Lite, and SWE-Bench-Verified \cite{sweverified}. 
Experimental results show that \method achieves state-of-the-art performance with minimal training overhead. 
With 14B-parameter model, we achieve resolved rates of \textbf{17.7\%} and \textbf{21.8\%} on \textbf{SWE-Bench-Lite} and \textbf{SWE-Bench-Verified}. 
This performance represents the best results among models with comparable parameters within open-source methods. 
Furthermore, \method achieves performance comparable to several commercial products with only hundreds of optimization training samples, highlighting its efficiency and practicality.  
Inspired by projects like OpenHands \cite{wang2024openhands}, we also build an agent-based software development platform using \method, which has already automated the creation of multiple trivial applications.
Human evaluations of these constructed applications demonstrate significant improvements in both effectiveness and user experience, highlighting the potential of \method to accelerate software automated development.

The key contributions of our work are listed below:

\begin{itemize}[leftmargin=10pt,itemsep=0pt,parsep=0pt]

\item We conduct a comprehensive analysis of existing code generation models, revealing a significant misalignment issue between existing training methods and the requirements from real-world software development agentic tasks. 

\item We propose a novel alignment framework specifically designed for real-world software engineering agentic workflows, namely \method. It employs a fine-grained alignment method utilizing MCTS for multi-step processes, and subsequently aligns with real-world requirements through preference optimization. 

\item We carry out experiments on three standard benchmarks for real-world software engineering agents, \ie, HumanEvalFix, SWE-Bench-Lite, and SWE-Bench-Verified, demonstrating \method's state-of-the-art performance with minimal training overhead.  
\end{itemize}  

\method enables code models to understand and use existing software engineering agentic workflows, making the ultimate sweet fruit of fully automated software development much more feasible.
Hopefully, this research could inspire subsequent researchers and contribute to the ultimate realization of the goal.

\section{Related Work}

\subsection{Large Language Models for Code}
Code generation, where models produce source code snippets from natural language (NL) descriptions, has been a hot-spot research area in recent years \cite{zhang2024codesurvey}.
LLMs have shown remarkable capabilities in this domain, which can be attributed to their extensive training on diverse datasets \cite{lozhkov2024starcoder}.
These models are often further fine-tuned through supervised fine-tuning (SFT) to maximize their coding potential and other abilities \cite{zhang2023instruction}.
Given the challenges associated with collecting high-quality training data for SFT, researchers have increasingly turned to self-instruct methods. In these techniques, the most powerful LLMs are employed to synthesize instruction data \cite{wang2022self,luo2023wizardcoder, wei2023magicoder}.
For instance, Evol-Instruct \cite{luo2023wizardcoder} makes use of complex prompts to improve the quality of the synthesized data.
OSS-instruct \cite{wei2023magicoder}, on the other hand, leverages real-world code snippets to enhance the relevance and practicality of the generated data.

The pre-training stage endows LLMs with excellent language and programming abilities, while the SFT stage enables them to follow instructions and complete a series of downstream tasks.
Researchers have thoroughly demonstrated that LLMs have achieved human-level performance on competitive programming benchmarks and beyond \cite{li2022competition, gpto3}, such as HumanEval \cite{chen2021evaluating} and LiveCodeBench \cite{jain2024livecodebench}.
To master the more complicated tasks that originate from real-world scenarios, post-training techniques are then introduced.

\subsection{Alignment Post-Training for Code Models}

While pre-training equips code models with foundational knowledge of programming and SFT improves their instruction following capabilities, post-training techniques adapt models to perform well in diverse and complex applications \cite{kumar2025llmposttrainingdeepdive}.
Recent research has increasingly highlighted the critical role of alignment training in shaping model performance and user experience, as evidenced by models such as ChatGPT and DeepSeek-R1 \cite{guo2025deepseekr1}.
In the realm of code, alignment training remains in its nascent stage.
Code-Optimise \cite{codeoptimise}, CodeDPO \cite{codedpo} and PLUM \cite{plum} construct post-training datasets from trivial programming problems, and align code models to perform greatly in such competitive programming tasks.
However, these methods concentrate on solving algorithmic problems from platforms such as LeetCode and Codeforces \cite{codeforces,leetcode}. While valuable, they do not fully capture the complexities inherent in real-world software development.
Beyond writing algorithmic code, practical software engineering demands a broader range of skills, \eg, navigating complex codebases, adapting and rewriting code snippets to specific contexts, and restructuring software architecture \cite{zhang2024codeagent, swebench}.
This gap between competitive programming and real-world software development underscores the need for more comprehensive alignment training methods tailored to practical software engineering agentic workflows.

\section{A Motivating Example}

\begin{figure}[t]
\centering
  \includegraphics[width=\columnwidth]{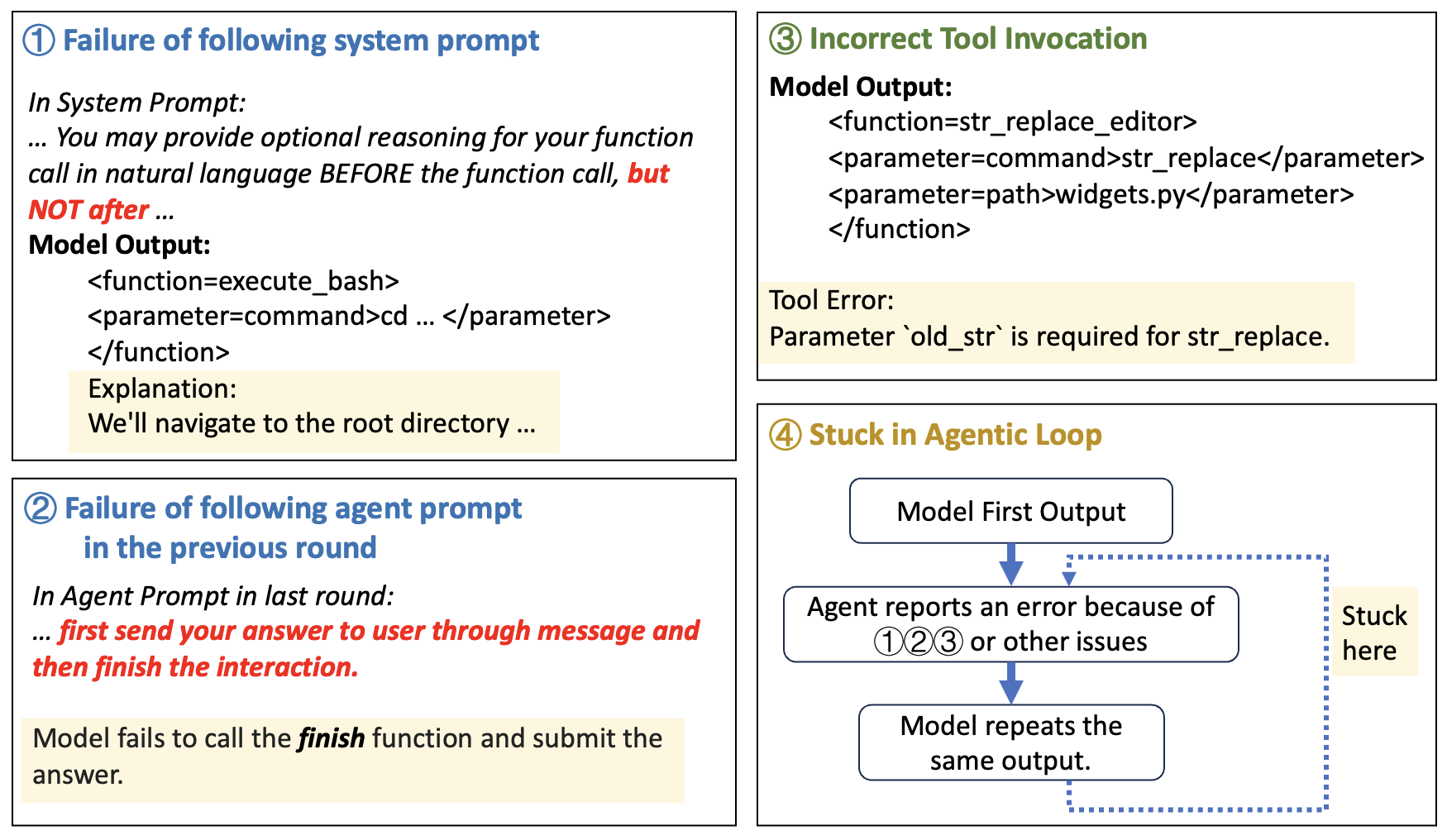}  
\caption{Observed cases of misalignment in existing code models and agentic frameworks.}
\label{fig:motivating}
\end{figure}

We present a motivating example to elaborate on the misalignment issue within current LLMs for code. Despite their great performance under algorithmic code generation, LLMs still struggle in benchmarks simulating real-world scenarios. 

\subsection{Code Agents \& Agentic Tasks}  

LLMs have demonstrated impressive performance in various code-related tasks, such as code completion, code summarization, and code generation \cite{allamanis2018survey}.
However, they can mostly be considered as standalone tasks, since LLMs are evaluated to solve trivial problems without requiring external knowledge or interacting with the environment.
Taking competitive programming as an example, the desired algorithmic code snippets are typically short and independent of other libraries or packages.

On the contrary, in the real-world, developers usually work with specific code environments to tackle different software engineering tasks \cite{zhang2024codeagent}.
They are supposed to extend functionalities based on existing contexts, locate and fix run-time errors, as well as review code commits from others, \etc
All these complex tasks require the ability to navigate large codebases, understand software architecture, and generate contextually appropriate code snippets.
It is clearly impractical and impossible to feed the entire codebase into LLMs, as this would exceed their maximum context and computational limitations.

To address real-world requirements, researchers have begun to explore agentic frameworks for handling complex tasks \cite{swebench, zhang2024codeagent}.
Agentic systems such as OpenHands \cite{wang2024openhands}, SWE-Agent \cite{yang2024swe}, CodeAgent \cite{zhang2024codeagent}, and AutoCodeRover \cite{zhang2024autocoderover} have demonstrated significant potential in addressing complex software engineering tasks.
SWE-Bench \cite{swebench} further provides a real-world software engineering benchmark by collecting GitHub issues and evaluating LLMs to address them (\eg, fixing bugs or adding new features for specific repositories).
These agentic solutions enable LLMs to autonomously explore and interact with the environment, with multiple pre-defined programming tools and workflow constraints.
For example, in the OpenHands framework \cite{wang2024openhands}, agents often need numerous interactive rounds to resolve a task.
Upon initialization, the LLM reads the task prompt ($p$) and generates its first action ($a_1$). Subsequently, in each round ($i$), the LLM interacts with the environment by invoking a tool using its output ($a_i$), and the environment responds with an observation ($o_i$) that reflects the changes resulting from the tool invocation.
The complete process of the LLM resolving a task can be characterized by the following trajectory:

\begin{equation}
    \text{Trajectory} = \langle p, a_1, o_1, a_2, o_2, \dots \rangle
\end{equation}

\noindent where the entire trajectory is typically very long, reflecting the complexity of real-world software engineering tasks.  
Furthermore, real-world code projects exhibit a high degree of diversity and context dependency. This complexity presents a significant challenge in constructing datasets and training LLM that can generalize across different complex scenarios.

\subsection{Failure Caused by Misalignment}  
\label{sec:failuremisalign}
After introducing agentic frameworks and defining trajectories, we conduct a preliminary experiment to show the misalignment issue of existing LLMs for code, and further analyze their failure cases.
Specifically, we adopt Qwen2.5-Coder-Instruct-14B \cite{qwencoder} as the code model and Openhands \cite{wang2024openhands} as the agentic framework.
We evaluate the constructed agent against the SWE-Bench-Lite benchmark \cite{swebench}.
To our shock, the agent can only resolve 3.7\% of problems within the benchmark.
After diving into the failure cases (as shown in Figure \ref{fig:motivating}), we find that one of the major causes of failures is that the base LLM can hardly follow the complicated instructions from the agentic framework.
Besides, the LLM often fails during tool invocation (\eg, select incorrect tools or use tool erroneously), and gets stuck in action loops (\eg, infinitely repeat one action).
We analyze these three main failure causes as below.

\paragraph{Failure of Instruction Following}  
The prompts in agentic frameworks are usually rather complicated. Due to the difficulty of the real-world tasks, the prompts often include additional information such as output constraints (the first issue in Fig. \ref{fig:motivating}), and workflow stages (the second issue in Fig. \ref{fig:motivating}), \etc
Although existing already-aligned code models perform well on standalone tasks, they frequently misinterpret such complex prompts or fail to execute multi-step instructions that require understanding context, navigating codebases, or invoking external tools.
This limitation highlights a critical gap between their objectives during training and the demands of practical software engineering tasks.  

\paragraph{Incorrect Tool Invocation}  
Another common failure mode is that LLMs invoke tools erroneously. \Eg, an LLM may select an incorrect tool from the available options or invoke a tool in an improper manner (such as using the wrong parameter in the third issue in Fig. \ref{fig:motivating}). 
Nevertheless, real-world software engineering tasks often require LLMs to invoke and interact with appropriate tools, such as text editors, linters, or search engines. 
Existing LLMs often make suboptimal choices during tool invocation, resulting in incomplete or incorrect solutions. This issue highlights the critical need for better alignment between LLMs and the tools they are intended to use. 

\paragraph{Stuck in Agentic Loop}  
In addition to the above two reasons, the agentic system has the tendency to get stuck in repetitive or unproductive loops.
\Eg, as shown in Fig. \ref{fig:motivating}, an LLM might repeatedly attempt the same action without making progress, or generate a sequence of actions irrelevant to the task. This behavior is particularly problematic in long trajectories, where the model’s inability to jump out of such loops often leads to task failure or excessive computational costs.

\section{\method}  

\begin{figure*}[t]
\centering
  \includegraphics[width=2\columnwidth]{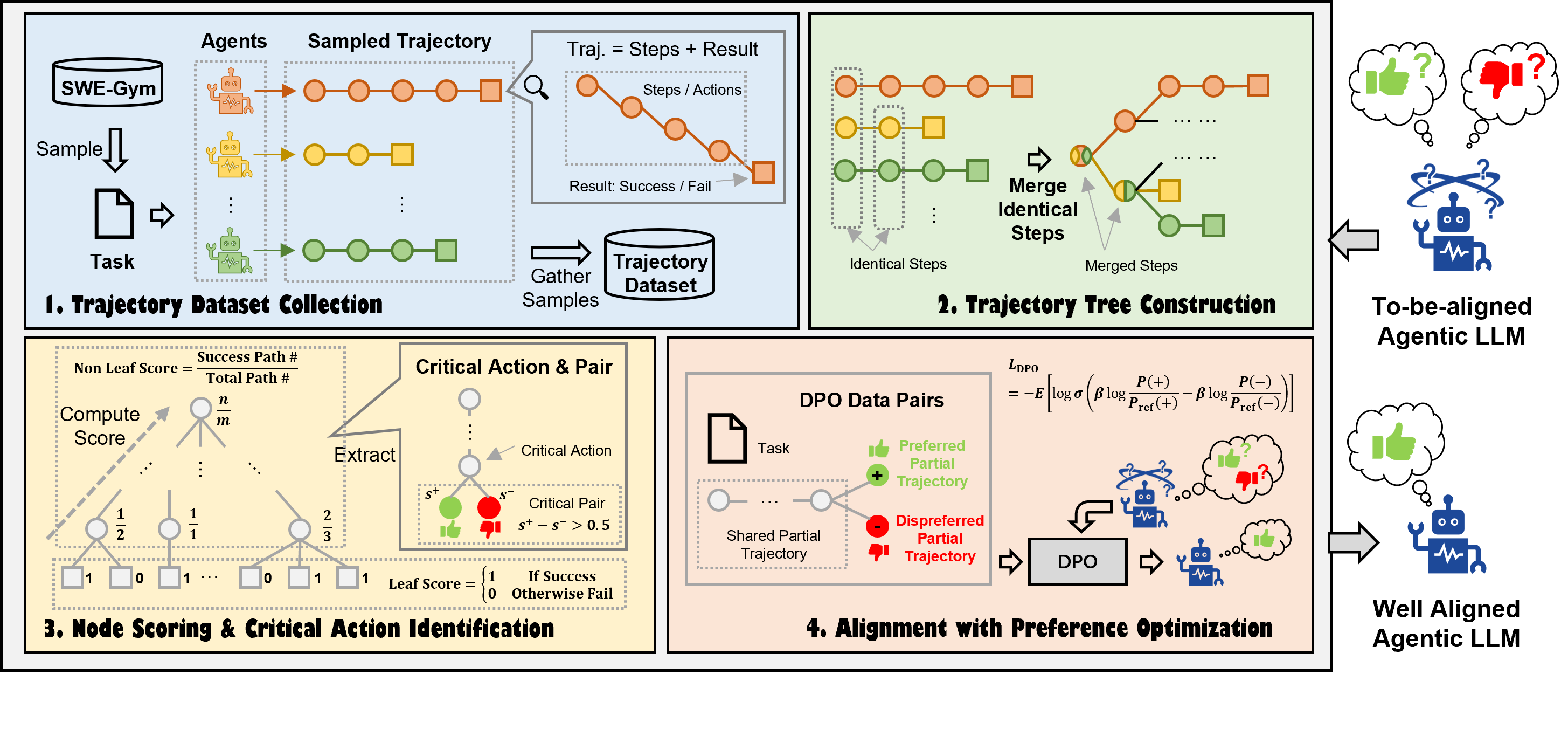}  
\caption{An overall pipeline of \method (best viewed in color). Our framework involves four steps: \ding{182} collecting agentic trajectory dataset from real-world software engineering environments, namely SWE-Gym \cite{swegym}, \ding{183} aggregating all sampled trajectories and constructing trajectory trees, \ding{184} scoring nodes within trajectory trees and extracting partial trajectory pairs with significant impact, and \ding{185} optimizing model preference with critical nodes. After these steps, we can finally obtain a well-aligned agentic LLM. Note that in step \ding{185}, it is usual to SFT the LLM before DPO as a warmup for the instruction following. To facilitate viewing, we do not include the SFT part in the pipeline figure.}
\label{fig:method}
\end{figure*}

We introduce \method as a novel alignment training framework designed to enhance code models for modern software engineering agentic systems.
Unlike traditional alignment solutions that focus on standalone algorithmic code generation tasks, \method aligns LLMs' core capabilities with specific agentic frameworks to better address real-world software development challenges.
\method is not only robust but also training-efficient, making it a feasible and practical option for applications.  

The framework of \method uses partial trajectory pairs to post-train LLMs. To collect these pairs with preference, we thoughtfully design a trajectory tree pipeline. Figure \ref{fig:method} elaborates the overall framework of \method.
\ding{182} \textbf{Trajectory Dataset Collection}.
We sample trajectories produced by agents constructed with top-performing LLMs and state-of-the-art agentic frameworks. Each trajectory consists of a sequence of actions and the final result. The trajectories are scored based on their corresponding resolved results ($1$ for success and $0$ for failure).
\ding{183} \textbf{Trajectory Tree Construction}. 
For each task, we gather all sampled trajectories to construct a trajectory tree. In this tree, identical action nodes (\eg, tool calls and model outputs) are merged, transforming individual agent decision chains into a unified tree structure. Such a trajectory tree can be considered an MCTS tree, where each path from the root to a leaf node refers to a plausible sample of a complete agent execution process.
\ding{184} \textbf{Node Scoring and Critical Action Identification}.
We design a scoring mechanism for nodes in the MCTS tree. Each intermediate node is scored based on the success rate of all paths in its subtree. After scoring every node, we identify actions that significantly impact the final outcome, which we define as critical agent actions.
\ding{185} \textbf{Alignment Training with Preference Optimization}.
we propose a specialized alignment training approach for agents. Focusing on the identified critical agent actions, we perform fine-grained preference optimization. We expect the aligned LLM to produce more ``good'' actions and less ``bad'' ones.
More specifically, ``good'' actions lead to a higher success rate while ``bad'' ones tend to fail.
We align the model with correct agent trajectories, thereby ensuring that it learns to prioritize effective decision-making paths. This approach achieves strong performance with minimal training data and computational cost.  

\subsection{Trajectory Dataset Collection}

To initiate model alignment, we follow existing work \cite{swegym}, where a training dataset with human-validated execution environments tailored for SWE-Bench-style tasks is provided. Based on this, we construct an initial trajectory dataset derived from high-performing models such as GPT-4o \cite{GPT-4o} and DeepSeek-v3 \cite{deepseekv3}. These models’ interactions within various software engineering agentic frameworks are collected to represent a series of decision-making steps taken by agents (\ie, invocations of tools).  

Each trajectory is paired with its corresponding final result, determined by whether the task succeeds (\ie, passes all test cases).
To avoid data leakage, we carefully exclude all repositories included in our evaluation benchmarks (SWE-Bench \cite{swebench} and HumanEvalFix \cite{HumanEvalFix}).
Every collected trajectory can then be formulated as below:

\begin{equation}
    \text{Trajectory} = \left\{\langle p, a_1, o_1, a_2, o_2, \dots \rangle, \text{res}\right\},\text{~where~res} \in \{1, 0\}
\end{equation}

\noindent where $\text{res}$ is a binary value, referring to the resolved result.
To ensure data integrity, we perform necessary de-duplication on the constructed trajectories. 

\subsection{Trajectory Tree Construction}

\method builds trajectory trees for the purpose of scoring and gathering nodes with impacts.
We observe frequent overlaps in the decision steps across different trajectories. This leads to the idea of merging identical steps to form a hierarchical trajectory tree structure, where all trajectories for a particular task are integrated into a unified structural representation.  

As shown in Figure \ref{fig:method}, for each task, we traverse the various trajectories and identify common prefixes across them. These shared prefixes serve as decision pathways that are merged into a single branch of the tree. By aligning identical steps across trajectories, we create a tree structure where \ding{182} each non-leaf node refers to an individual decision step (\eg, a tool invocation or an action taken by the agent) and \ding{183} each leaf node represents the final outcomes of the trajectory (\ie, success or failure of the task).

With the trajectory tree, \method organizes agent action chains into a hierarchical structure, allowing effective and efficient node scoring and critical action analysis across trajectories.
A filtration procedure is further introduced to eliminate low-quality trajectories that contribute little to the model's performance.
There are two main filtering rules:

\begin{itemize}[leftmargin=10pt,itemsep=0pt,parsep=0pt]
\item Trajectories repeating the same actions for multiple steps without progress are eliminated, which often leads to the stuck-in-loop issue;
\item Trajectories that do not overlap in any meaningful way with others are removed, which usually suggests irrelevant outliers.
\end{itemize}
\noindent This filtering procedure ensures that the dataset maintains a high standard of quality and relevance, allowing the trajectory tree to focus on representative pathways. 

\subsection{Node Scoring \& Critical Action Extraction}

\begin{algorithm}[t]
\caption{Node Scoring and Critical Action Identification.}
\label{alg:node_scoring}
\begin{algorithmic}[1]
\Procedure{ComputeNodeScores}{traj\_tree}
    \State $\text{scores, critical\_actions, critical\_pairs} \gets \{\}, \{\}, \{\}$ \Comment{Init} % \Comment{Stores success rates for nodes.}
    % \State Initialize $critical\_actions \gets \{\}$ % \Comment{Tracks critical actions.}
    % \State Initialize $critical\_pairs \gets \{\}$ % \Comment{Tracks preference pairs.}
    \For{each $\text{node} \in \text{PostOrderTraverse}(\text{traj\_tree})$}
        \If{$\text{node}$ is a leaf}
            \State $\text{scores}[\text{node}] \gets 0 \text{~or~} 1$ \Comment{Determined by the result}
            % \State $\text{ based on } \text{ task execution output of entire trajectory}$
        \Else
            \State $\text{scores}[\text{node}] \gets \frac{\text{Successful path \# from node}}{\text{Total path \# from node}}$
        \EndIf
        % \State $scores[node] \gets success\_rate$
        \If{$\text{node}$ has two children with $\text{score\_diff} > 0.5$}
            \State $\text{critical\_actions}\gets\text{critical\_actions}\cup\{\text{node}\}$
            % \State Mark $node$ as $critical\_actions$
            \State $\text{critical\_pair}\gets$Paired partial trajs with preference
            \State $\text{critical\_pairs}\gets\text{critical\_pairs}\cup\{\text{critical\_pair}\}$
            %\State Add $(correct\_partial\_traj, wrong\_partial\_traj)$ to $critical\_pairs$
        \EndIf
    \EndFor
    \State \textbf{return} $\text{scores}$, $\text{critical\_actions}$, $\text{critical\_pairs}$
\EndProcedure
\end{algorithmic}
\end{algorithm}

After constructing the trajectory trees, \method is allowed to identify critical pathways and actions, which significantly influence the final results. Please refer to Algorithm \ref{alg:node_scoring}.
To achieve so, we develop a \textbf{node scoring mechanism} based on the success rates of subtrees emanating from each node. The scoring process is outlined as follows.
\ding{182} For each leaf node, its score is assigned to $0$ or $1$ according to the final result.
\ding{183} The score for a non-leaf node is computed by propagation in a bottom-up manner -- it is the fraction of success path numbers over all paths in the node's subtree.
In general, the scores reflect the plausible success rate of all possible paths extending from this node.

With the assistance of node scores, \method identifies critical actions in the trajectory tree.
If there is a significant difference in the scores of two non-leaf nodes under the same parent node, it indicates that these two non-leaf nodes correspond to a set of critical actions, meaning they lead to vastly different outcomes (one for a high success rate and the other for a high failure rate).
In our experimental setup, the threshold of score differences is set to $0.5$.
During alignment training, we aim to train code models using these critical action pairs to ensure that the models consistently produce higher score actions -- that is, actions associated with higher success rates.
Please refer to Table \ref{table:trajectory_stats} for statistics of the collected trajectory dataset.

\begin{table}[t]
\centering
\caption{Statistics of Trajectories and Trajectory Trees}
\label{table:trajectory_stats}
\begin{tabular}{clc}
\toprule
& \textbf{Metric} & \textbf{Value} \\
\midrule
%\textbf{Trajectory Collection} \\
\multirow{5}{*}{\makecell{\textbf{Collected}\\ \textbf{Trajectory}}}
& Instance \# & 186 \\
& Trajectory \# (After Filtration) & 3,168 \\
& \quad -- Successful Trajectory \# & 677 \\
& \quad -- Wrong Trajectory \# & 2,491 \\
& Avg. Token Len. & $\sim$19,868 \\
\midrule
%\textbf{Trajectory Tree} \\
\multirow{2}{*}{\makecell{\textbf{Trajectory}\\\textbf{Tree}}}
& Avg. Path Len. & $\sim$41 \\
& Critical Pair \# & \textbf{983} \\
\bottomrule
\end{tabular}
\end{table}

\subsection{Alignment with Preference Optimization}  
\label{sec:trainingloss}

With the collected trajectory dataset and the critical actions, \method aligns open-source LLMs to high-score actions.
Based on the trajectory dataset constructed above, and the critical action nodes identified through the trajectory tree scoring mechanism, we perform subsequent \textbf{alignment training} on open-source models. The entire post-training process consists of two phases: \ding{182} SFT with correct trajectories, and \ding{183} Direct Preference Optimization (DPO) with critical action pairs.

\paragraph{SFT with Correct Trajectories}  
Before actual alignment, it is crucial to ensure that the target LLM can generate valid and appropriate trajectories, and therefore SFT is needed. Given a collected correct trajectory data sample ($\langle p, a_1, o_1, a_2, o_2, \dots \rangle$), the target LLM should learn the actions ($a_1,a_2,\cdots$).
Thus, during SFT, we mask out the loss for the environment observations ($o_1,o_2,\cdots$), and the loss is formulated as:

\begin{equation}
\mathcal{L}_{\text{SFT}} = -\mathbb{E}_{\text{traj} \in \mathcal{D}} \left[ \sum_{t=1}^{T} \log P(a_t \mid p, a_1,o_1, \dots,a_{t-1},o_{t-1}) \right],
\end{equation}

\noindent where $T$ is the total number of actions in the trajectory, and \(\mathcal{D}\) is the dataset of all correct trajectories.

\paragraph{Fine-grained DPO with Critical Action Pairs}
A major challenge of SFT is that most trajectories appear similar, with only a few critical actions making a significant difference.
After SFT, the target LLM can generate valid actions in trajectories but requires further alignment to produce "good" actions (\ie, those with higher success rates).
The target model is fundamentally limited in distinguishing and optimizing for these critical actions.
To tackle this challenge, we adopt DPO \cite{rafailov2024direct}, which introduces a preference-based mechanism that extends beyond maximizing the likelihood of ground truth data.
Previous studies have shown the promise of DPO in reasoning tasks such as mathematics \cite{lai2024step, wu2024self, codedpo}.
By utilizing pairs of preferred responses ($y^+$) and dispreferred responses ($y^-$), DPO aligns the model with human preferences, thereby enhancing its effectiveness for tasks that require nuanced decision-making.
In \method, we focus on the critical agentic actions. As introduced before, the decision points which may significantly influence the final outcome are collected. Each critical pair contains a correct partial trajectory and a wrong partial trajectory, which share the same action prefix, formulated as below:

\begin{equation}
    \text{Critical Pair}=\langle a_i^+,a_i^-\mid p,a_1,o_1,\cdots,a_{i-1},o_{i-1} \rangle
\end{equation}

\noindent where $\langle p,a_1,o_1,\cdots,a_i^+ \rangle$ refers to the preferred ``good'' partial trajectory and $\langle p,a_1,o_1,\cdots,a_i^- \rangle$ refers to the dispreferred ``bad'' partial trajectory.
Remember that the score difference threshold is $0.5$, Under the same partial trajectory condition ($\langle p,a_1,o_1,\cdots,a_{i-1},o_{i-1} \rangle$), the preferred action ($a_i^+$) leads to much higher success rate, making it critical. DPO can effectively help the code model to learn such preference, and the loss is formally defined as below:

\begin{equation}
\mathcal{L}_{\text{DPO}} = -\mathbb{E}_{\text{traj} \in \mathcal{D}} \Bigg[ \log \sigma \Bigg(\beta \log \frac{P(a_i^+)}{P_{\text{ref}}(a_i^+)} - \beta \log \frac{P(a_i^-)}{P_{\text{ref}}(a_i^-)} \Bigg) \Bigg] \label{eq:dpo_loss}
\end{equation}

\noindent where \(P(\cdot)\) is the policy model's likelihood for making decisions, and \(P_{\text{ref}}(\cdot)\) is the reference model's likelihood. Note that all conditions in the formulation are neglected for a better view (\eg, $P(a_i^+)$ in Eq. \ref{eq:dpo_loss} should have been $P(a_i^+|p,a_1,o_1,\cdots,a_{i-1},o_{i-1})$).

\section{Experimental Setups}  

To evaluate the effectiveness of \method, we conduct a comprehensive large-scale study. We aim to answer the following research questions (RQs):

\noindent\textbf{RQ1: Does \method improve the correctness of the generated code on non-trivial real-world benchmarks?}  
We apply \method to variants of SWE-Bench, \ie, SWE-Bench-Lite \cite{swebench} and SWE-Bench-Verified \cite{sweverified}. Both benchmarks assess LLM's capability to handle non-trivial real-world software engineering tasks, such as bug fixing and new feature implementation.

\noindent\textbf{RQ2: Does \method improve the agent capability on more general code-related tasks?}  
According to the motivating example in Section \ref{sec:failuremisalign}, LLMs often struggle to follow complex instructions or have trouble in correctly utilizing tools in software engineering agent workflows. To demonstrate that \method enhances agent capability with minimal training overhead, we evaluate its performance on a general code-related benchmark, namely HumanEvalFix \cite{HumanEvalFix}.

\noindent\textbf{RQ3: What is the impact of the fine-grained preference optimization method on agent capability in \method?}  
The data construction and the loss formulation may have an influence upon \method. To demonstrate the effective design of \method, we conduct ablation studies by comparing different training setups, including variations in dataset construction and key components of our training loss formulation.  

\noindent\textbf{RQ4: How does data scaling influence the performance of \method?}  
Data scaling is a crucial factor for modern LLMs, as it reflects the efficiency of the training schema.
We explore the impact of data scaling on \method by varying the amount of training data. This analysis shows how data size affects the ability of LLMs to handle real-world software engineering tasks.  

\noindent\textbf{RQ5: How does \method improve the user experience in real-world software development scenarios?}  
In addition to quantitative analysis, we also conduct human evaluations to assess the effectiveness of \method in enhancing user experience. We construct multiple applications using \method, and study improvements in user experience, in order to demonstrate the \method framework's practical utility.  

\subsection{Benchmarks}

We conduct our main experiments upon three benchmarks, \ie SWE-Bench-Lite \cite{swebench}, SWE-Bench-Verified \cite{sweverified}, and HumanEvalFix \cite{HumanEvalFix}.
\ding{182} SWE-Bench-Lite \cite{swebench} is a subset of SWE-Bench. The original SWE-Bench is a large-scale benchmark assessing the LLMs with thousands real-world GitHub issues, spanning 12 popular Python repositories. SWE-Bench closely simulates real-world software engineering scenarios, where models are tasked with fixing bugs or implementing new features. To better accommodate functional bug-fixing objectives, SWE-Bench-Lite is curated, containing 300 test instances and 23 development instances sampled from SWE-Bench. The instances in SWE-Bench-Lite are more self-contained, focusing on functional bug-fixing tasks. 
\ding{183} SWE-Bench-Verified \cite{sweverified} is another variation of SWE-Bench. OpenAI employs human beings to thoroughly validate instances in SWE-Bench, creating a human-validated dataset. SWE-Bench-Verified contains 500 verified test instances in total.
\ding{184} In addition to variants of SWE-Bench, we also evaluate \method on HumanEvalFix \cite{HumanEvalFix}, which is a general debugging benchmark. The benchmark introduces bugs into each of the 164 HumanEval \cite{chen2021evaluating} solutions, and requires the model to correct functions with subtle bugs, accompanied by relevant unit tests. 

SWE-Bench-Lite and SWE-Bench-Verified are widely adopted benchmarks for LLM-based software engineering agents. Following prior works, we evaluate \method on these two benchmarks for a consistent and fair comparison.
In contrast to the SWE-Bench family, HumanEvalFix is a bit simpler but much more generic, making it a complementary testbed to assess \method's ability to improve instruction-following capabilities within agentic frameworks. 
Note that during data processing in the \method pipeline, we exclude all the repositories and code snippets used by these three benchmarks to prevent data contamination.

\subsection{Metrics}

For SWE-Bench family, we employ the resolved rate as the primary evaluation metric, which measures the percentage of successfully resolved problems. In addition to the primary metric, we also include the following metrics for more detailed analysis \cite{wang2024openhands}:  \ding{182} empty patch rate, which estimates the percentage of trajectories where no code in the repository is edited, and \ding{183} stuck-in-loop rate, which assesses the percentage of trajectories where the agent gets stuck in an agentic loop by repeating the exact same action for 3 rounds.  
The evaluations of the subject LLMs are limited by either a maximum of 30 interaction turns or the model's 32k context window length, whichever is reached first.

For HumanEvalFix, we use the pass rate on the provided unit tests as the primary evaluation metric. As a complementary measurement, we also record the invalid patch rate, which indicates the percentage of trajectories in which the generated patch is incomplete or empty. Both metrics evaluate the agent's instruction-following capabilities under \method.

\subsection{Baselines}

We evaluate widely adopted LLMs in the realm of code generation, focusing on Qwen-2.5-Coder-Instruct-7B and Qwen-2.5-Coder-Instruct-14B \cite{qwencoder}. These two models are post-trained from Qwen-2.5-Coder using large-scale, million-level code-related datasets with SFT and DPO. They represent the state-of-the-art LLMs, specialized for coding, within their respective parameter size categories.

We adopt the OpenHands agentic framework \cite{wang2024openhands} to construct a training dataset and conduct alignment training for \method. OpenHands is a general-purpose agentic framework, which utilizes the CodeActAgent strategy \cite{codeactagent} for decision-making regarding both tool usage and reasoning steps.  

\subsection{Training \& Inference Setups}

We leverage powerful general LLMs like DeepSeek-v3 to build agents for data construction in \method. The constructed dataset focuses exclusively on Python, given its widespread usage and relevance for software engineering tasks.

The maximum training epoch number is set to 10, and the best-performing checkpoint is selected based on the lowest validation loss. Learning rate is initialized to $3 \times 10^{-6}$, equipped with a linear scheduler and a warm-up phase at the start of training. 
Greedy search is adopted during inference, as it ensures deterministic and consistent results for evaluation.  

\section{Experimental Results}

\subsection{SWE-Bench Family (RQ1)}

To answer \textbf{RQ1}, we first compare the results of \method with the baseline LLMs. The comparison evaluates the improvements \method achieves over widely used models (\ie, Qwen-2.5-Coder-Instruct-7B and Qwen-2.5-Coder-Instruct-14B) in terms of resolved rates on SWE-Bench-Lite and SWE-Bench-Verified. Detailed results are presented in Table \ref{tab:baseline_comparison_lite}.  

Results show that \method achieves significant relative improvements over baseline LLMs. 
The baseline LLMs, including the 14B parameter variant, struggle to effectively utilize the agentic framework to handle SWE-Bench tasks due to the misalignment issue, as discussed in Section \ref{sec:failuremisalign}.  
With alignment achieved through a relatively small training cost across less than a thousand samples, \method successfully resolves the misalignment between existing code models and agentic frameworks. \method significantly enhances the target LLMs' instruction following and tool utilization capabilities within the agentic framework.  
Notably, \method demonstrates remarkable improvements in addressing issues such as empty patches and stuck-in-loop, highlighting the efficiency of our alignment method in tackling real-world software engineering challenges.

\begin{table}[t]
\centering
\caption{Performance Comparisons on SWE-Bench-Lite and SWE-Bench-Verified. ``Empty'', ``Stuck'', ``Resolve'' denote empty patch, stuck-in-loop, and resolved reates, respectively. $\Delta$ denotes the relative improvement between the baseline and \method. "Qwen-7B" is an abbreviation for Qwen2.5-Coder-Instruct-7B, and "Qwen-14" follows the similar abbreviation.}
\label{tab:baseline_comparison_lite}
%\resizebox{\linewidth}{!}{
\begin{tabular}{lcccc}
\toprule
Model & {\footnotesize Empty(\%)~$\downarrow$} & {\footnotesize Stuck(\%)~$\downarrow$} & {\footnotesize Resolve(\%) ~$\uparrow$} & {\footnotesize $\Delta(\%)$~$\uparrow$} \\
\midrule
\multicolumn{4}{l}{\textbf{\textit{SWE-Bench-Lite}}\quad\textbf{\textit{Instance \# = 300}}} \\
\midrule
Qwen-7B & 53.0 & 45.3 & 1.0 & - \\
\method-7B & 22.0 & 24.0 & 15.0 & +1400.0 \\
\addlinespace
Qwen-14B & 50.3 & 35.0 & 3.7 & - \\
\method-14B & 17.0 & 22.0 & 17.7 & +378.3 \\
\toprule
\multicolumn{4}{l}{\textbf{\textit{SWE-Bench-Verified}}\quad\textbf{\textit{Instance \# = 500}}} \\
\midrule
Qwen-7B & 49.4 & 39.0 & 0.8 & - \\
\method-7B & 25.2 & 19.2 & 18.6 & +2225.0 \\
\addlinespace
Qwen-14B & 52.0 & 27.8 & 2.8 & - \\
\method-14B & 22.8 & 15.6 & 21.8 & +678.5 \\
\bottomrule
\end{tabular}
%}
\end{table}

We further compare the performance of \method with other state-of-the-art agent-based methods, including both commercial applications and open-source approaches. As shown in Table \ref{table:sota_comparison}, \method consistently outperforms many of these state-of-the-art methods in terms of task resolved rates on both SWE-Bench-Lite and SWE-Bench-Verified.  

Notably, \method, using the same OpenHands agentic framework and a 14B LLM, achieves the comparable performance of GPT-4o, which is a commercial closed-source application. Considering the intrinsic complexity of SWE-Bench tasks, the results are particularly surprising, highlighting the effectiveness of \method.  
On the other hand, among listed open-source methods, \method achieves the best results among LLMs with comparable parameter size, only being outperformed by a few existing 72B models. 
Through this comparison, we highlight the remarkable performance of the proposed method, which is achieved with a small training cost in this highly competitive landscape.

\begin{table}[t]
\centering
\caption{Comparisons with State-of-the-Art Agent Methods.}
\label{table:sota_comparison}
%\resizebox{\linewidth}{!}{
\begin{tabular}{llcc}
\toprule
\multirow{2}{*}{Method} & \multirow{2}{*}{\makecell{Model /\\Params}}  & \multicolumn{2}{c}{SWE-Bench} \\
    \cmidrule{3-4}
&& Verified (\%) & Lite (\%) \\
\midrule
\multicolumn{4}{l}{\textbf{Closed-source Methods}} \\
SWE-agent \cite{yang2024swe} & \makecell[l]{\footnotesize{Claude-3}\\\footnotesize{-Opus}} & 7.0 & 4.3 \\
    \cmidrule{2-2}
RAG \cite{swebench} & \multirow{2}{*}{GPT-4} & 2.8 & 2.7 \\
RepoGraph \cite{ouyang2024repograph} & & - & 19.0 \\
    \cmidrule{2-2}
OpenHands \cite{wang2024openhands} & \multirow{4}{*}{GPT-4o} & - & 21.0 \\
AutoCodeRover \cite{zhang2024autocoderover} & & - & 30.7 \\
SWE-SynInfer \cite{ma2024lingma} & & - & 22.0 \\
Agentless \cite{xia2024agentless} & & 31.8 & 20.7 \\
    \cmidrule{2-2}
Moatless Tools \cite{moatless} & \multirow{2}{*}{\makecell[l]{\footnotesize{Claude-3.5}\\\footnotesize{-Sonnet}}} & 33.6 & 23.0 \\
OpenHands \cite{wang2024openhands} & & 50.8 & 40.7 \\
\toprule
\multicolumn{4}{l}{\textbf{Open-source Methods}} \\
SWE-Llama RAG \cite{swebench} & 13B & 1.2 & 1.0 \\
AutoCodeRover \cite{zhang2024autocoderover,liu2024codexgraph} & 72B & - & 9.3 \\
SWE-Gym  \cite{swegym} & 7B & 10.6 & 10.0 \\
SWE-Gym  \cite{swegym} & 14B & 16.4 & 12.7  \\
SWE-Gym \cite{swegym} & 32B & 20.6  & 15.3 \\
Lingma SWE-GPT \cite{ma2024lingma} & 72B & 30.2 & 22.0 \\
SWE-Fixer \cite{xie2025swe} & 72B & 32.8 & 24.7 \\
\midrule
\method (Ours) & 7B & 18.6 &  15.0  \\
\method (Ours) & 14B & 21.8 & 17.7  \\
\bottomrule
\end{tabular}
%}
\end{table}

\subsection{HumanEvalFix (RQ2)}

To evaluate whether \method enhances an LLM's ability to follow complex instructions or correctly utilize tools in software engineering agentic workflows, we conduct experiments on HumanEvalFix, which is a relatively simpler but broader and more popular benchmark. The results are listed in Table \ref{table:HumanEvalFix_results}.
It is important to note that the dataset constructed by \method does not include any data related to HumanEvalFix. Additionally, the data styles and formats are significantly different between the two. This ensures that there is no risk of data leakage.
We would like to elaborate that \method exhibits generalizability and adaptability in \textbf{RQ2}.

Results highlight that \method significantly improves task success rates and reduces the invalid patch rate compared to baseline LLMs within the agentic framework. 
We also evaluate the performance of directly using the model to solve the task without any agentic framework. 
For the baseline Qwen models, there is a clear performance gap between using and not using the agentic framework.
In contrast, our proposed \method demonstrates robustness across both forms of workflow, achieving high performance either with or without the agentic framework.
Interestingly, we observe that incorporating the agentic framework further enhances accuracy for \method. This improvement can be attributed to the agentic framework’s provision of appropriate tools (such as executor), which facilitate code-related task resolution.  
These findings demonstrate that a well-aligned LLM for code obtained by \method, in combination with a compatible agentic framework, can achieve improved generalization across a broader range of code-related tasks.

\begin{table}[t]
\centering
\caption{Performance Comparisons on HumanEvalFix. ``Invalid'' denotes the invalid patch rate within the agentic mode. "Qwen-7B" is an abbreviation for Qwen2.5-Coder-Instruct-7B, and "Qwen-14" follows the similar abbreviation.
}
\label{table:HumanEvalFix_results}
% \resizebox{0.8\linewidth}{!}{
\begin{tabular}{lccc}
\toprule
{{Model}} & {{Agent}} & {{Pass@1 (\%) $\uparrow$}} & {{Invalid (\%) $\downarrow$}} \\
\midrule
\multirow{2}{*}{Qwen-7B} & w/o & 46.9 & - \\
                           & w/    & 18.9 & 60.4 \\
\addlinespace
\multirow{2}{*}{Qwen-14B} & w/o & 54.3 & - \\
                           & w/   & 31.1 & 43.3 \\
\midrule
\multirow{2}{*}{\method-7B} & w/o & 46.3 & - \\
              & w/  & 55.5 & 13.4  \\
\addlinespace
\multirow{2}{*}{\method-14B} & w/o & 52.4 & - \\
               & w/   & 62.8 & 10.4 \\
\bottomrule
\end{tabular}
% }
\end{table}

\subsection{Ablation Studies (RQ3)}

\begin{table}[t]
\centering
\caption{Ablation Studies on SWE-Bench-Lite.}
\label{table:ablation_results}
%\resizebox{\linewidth}{!}{
\begin{tabular}{lcc}
\toprule
{Method} & {Empty(\%)~$\downarrow$} & {Resolve(\%)~$\uparrow$} \\
\midrule
Qwen-14B (w/o all) & 50.3 & 3.7 \\
\method-14B (w/ all) & 17.0 & 17.7 \\
\midrule
\multicolumn{3}{l}{\textbf{Ablations on Training Stage}} \\
 w/ SFT, w/o Fine-grained DPO & 22.0 & 13.0 \\
 w/o SFT, w/ Fine-grained DPO & 24.3 & 10.7 \\
\midrule
\multicolumn{3}{l}{\textbf{Ablations on Fine-grained Score}} \\
 w/o Fine-grained Score & 41.3 & 5.3 \\
\bottomrule
\end{tabular}
%}
\end{table}

To answer \textbf{RQ3}, we evaluate the improvement at each training stage introduced in Section \ref{sec:trainingloss}. Specifically, we analyze the effects of SFT and fine-grained DPO on final performance. 
Additionally, we assess the impact of \method’s carefully designed components, including the computation of fine-grained scores in our DPO setting, as shown in Table \ref{table:ablation_results}.

Results show that the DPO stage can further enhance the model's instruction-following capabilities within agentic frameworks, building upon the foundation provided by the SFT stage. Both stages are crucial for achieving the final improvements in performance.  
Besides the training stages, the fine-grained scoring mechanism in our DPO setting plays a critical role. When critical action identification is omitted and DPO is applied directly to positive and negative samples (the entire trajectories that succeed or fail) of the same task, the results degrade significantly.  
This fine-grained preference optimization contributes substantially to task success rates by aligning LLMs' critical decision-making at key steps. These alignments on critical actions have a profound impact on the model's final performance.

\subsection{Data Scaling Law for \method (RQ4)}

\begin{table}[t]
\centering
\caption{Data Scaling for \method on SWE-Bench-Lite. The percentage values indicates the data ratio utilized for alignment training.}
\label{table:data_scaling_results}
% \resizebox{0.8\linewidth}{!}{
\begin{tabular}{lcc}
\toprule
{Method} & {Empty(\%) $\downarrow$} & {Resolve(\%) $\uparrow$} \\
\midrule
\multicolumn{3}{l}{\textbf{Data Scaling}} \\
Qwen-14B (0\%) & 50.3 & 3.7 \\
25\% & 35.0 & 9.0 \\
50\% & 28.7 & 13.7 \\
75\% & 22.0 & 16.0 \\
\method-14B (100\%) & 17.0 & 17.7 \\
\midrule
\multicolumn{3}{l}{\textbf{Iterative DPO in Sec. \ref{sec:onlinerl}}} \\
+ 50\% on-policy dataset & 17.0 & 16.3 \\
\bottomrule
\end{tabular}
% }
\end{table}

To address \textbf{RQ4}, we investigate the impact of scaling training data on the performance of \method.
We evaluate \method using training data proportions of 25\%, 50\%, and 75\%, and analyze their respective impacts on model performance. As listed in Table \ref{table:data_scaling_results}, training with more data examples consistently enhances the performance of the target LLM.

However, further scaling is constrained due to limitations such as extended execution time and the resource costs associated with constructing larger-scale datasets. For detailed discussions on time and efficiency, please refer to Section \ref{sec:timeefficiency}.
In our experiments, we balance performance gains with training costs, ensuring that improvements are achieved cost-effectively. Future work will explore efficient data scaling methods to investigate \method’s performance upper bounds under extreme training budgets.

\subsection{User Experience (RQ5)}

To answer \textbf{RQ5}, we follow the application demo settings of OpenHands \cite{wang2024openhands} and manually design five real-world development tasks: \ding{182} a to-do list web application, \ding{183} a snake game, \ding{184} a weather forecasting application, \ding{185} a Hacker News query application, and \ding{186} a personalized homepage.
For each task, we employ both baseline models (\ie, Qwen-based models) and \method within the OpenHands framework to automate development. 
Five independent volunteer evaluators, each with one year or more software developing experience, were invited to assess the results from three distinct perspectives: functionality completeness, code quality (\eg, readability and adherence to coding standards), and aesthetic appeal.
Additionally, we invited all evaluators to rank the five tasks by difficulty. The ranking results, along with corresponding human evaluation metrics, are summarized in Table \ref{table:user_experience}.

\begin{figure}[t]
\centering
  \includegraphics[width=0.8\columnwidth]{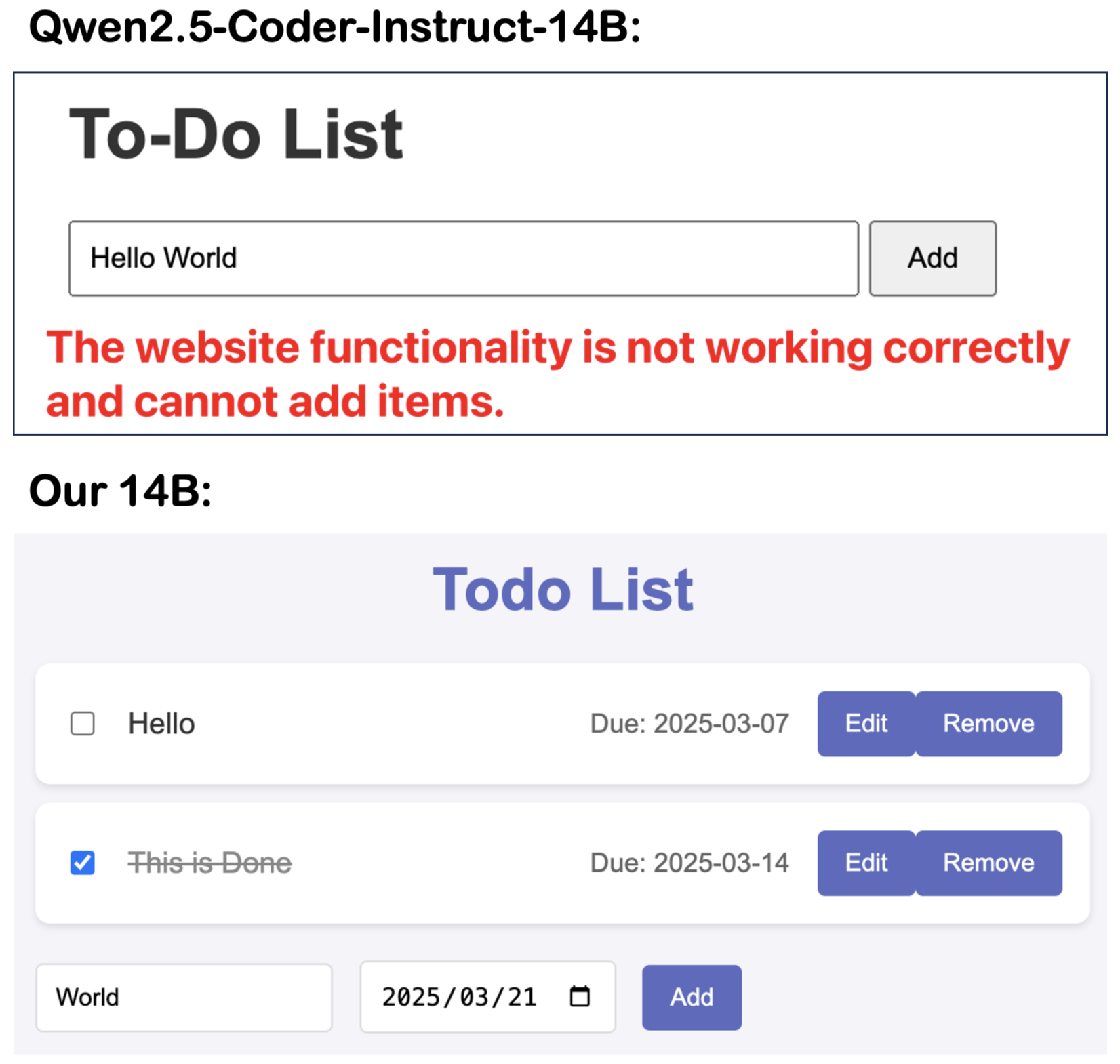}  
\caption{Case study on creating a to-do list web application with OpenHands agentic framework compared with \textbf{Qwen2.5-Coder-Instruct-14B} and \textbf{\method-14B}.}
\label{fig:casestudy}
\end{figure}

Results in Table \ref{table:user_experience} show that \method significantly enhances user experience in real-world development scenarios and demonstrates its practical utility.
A to-do list case is presented in Figure \ref{fig:casestudy}, where Qwen2.5-Coder-Instruct-14B struggles to produce a fully functional application, while \method successfully completes all required features perfectly\footnote{Please refer to our open-source reproduction package \cite{casestudy} for other cases.}.   
Upon further analyzing the agent execution chains, we find that the Qwen2.5 model frequently encounters issues such as incorrect tool invocation and failure to follow instructions. These errors cause the agentic framework to spend most of its time handling exceptions and recovering from erroneous steps, significantly hindering task completion. 
In contrast, \method addresses these shortcomings effectively. With minimal training overhead, \method achieves strong alignment with real-world software engineering scenarios, greatly improving the usability of open-source models in complex software development tasks and agentic frameworks.
\begin{table}[t]
\centering
\caption{User Experience Evaluation on Real-World Software Engineering Tasks. Results compare \textbf{Qwen2.5-Coder-Instruct-14B} and \textbf{\method-14B}. Each task is scored by five developers on a scale from 1 to 5, where 5 represents the best, and the average score is calculated. Tasks are ordered from top to bottom based on difficulty, and ranked from simplest to most complex according to evaluator assessments.}
\label{table:user_experience}
%\resizebox{\linewidth}{!}{
\begin{tabular}{llccc}
\toprule
{Task} & {Model} & {Func.} & {Quality} & {Aesthetic Appeal} \\
\midrule

\multirow{2}{*}{Homepage} & Qwen & 1.7 & 3.3 & 2.7 \\
& \method & 3.0 & 3.7 & 3.0 \\

\addlinespace
\multirow{2}{*}{Weather} & Qwen & 1.0 & 2.0 & 1.0 \\
& \method & 3.0 & 3.0 & 2.3 \\

\addlinespace
\multirow{2}{*}{News} & Qwen & 4.0 & 4.3 & 3.3 \\
& \method & 4.3 & 4.3 & 4.0 \\

\addlinespace
\multirow{2}{*}{To-do} & Qwen & 1.3 & 3.0 & 1.7 \\
& \method & 3.7 & 3.3 & 4.3 \\

\addlinespace
\multirow{2}{*}{Snake} & Qwen & 1.0 & 1.0 & 1.3 \\
& \method & 1.7 & 3.0 & 2.3 \\

\midrule
\addlinespace
\multirow{2}{*}{Avg.} & Qwen & 1.8 & 2.7 & 2.0 \\
& \method & 3.1 & 3.5 & 3.2 \\

\bottomrule
\end{tabular}
%}
\end{table}

\section{Discussions \& Future Work}

\subsection{Generalization across Agentic Frameworks}

Just as experienced human programmers skillfully utilize their preferred Integrated Development Environments (IDEs), LLMs optimized for specific agents can effectively leverage tools to address complex problems. 
As demonstrated in our experiments in \textbf{RQ1}, even relatively small-scale LLMs, when subjected to framework-specific alignment, can achieve a high degree of proficiency in agentic frameworks, thereby unlocking new possibilities for automation and broader applications.

However, a key question remains -- can LLMs trained on one specific agentic framework generalize to others? 
To address this, we conduct experiments to evaluate the generalization capability of \method. Specifically, a model optimized for the OpenHands framework using \method is evaluated on the AutoCodeRover framework \cite{zhang2024autocoderover}. While OpenHands is a flexible and general-purpose agentic framework capable of dynamic tool calls, AutoCodeRover relies on domain-specific, pre-defined workflows that provide the model with different tools at various stages to solve corresponding subproblems. We still employ Qwen2.5-Coder-Instruct-14B and \method-14B to evaluate their performance. In addition, we also compare with the close-sourced powerful LLM, GPT-4o \cite{GPT-4o}.

The experimental results, presented in Table \ref{table:autocoderover}, show that while the aligned LLM, transferred from OpenHands to AutoCodeRover, achieves reasonable performance improvements over the unaligned baseline, its performance remains significantly lower than that of powerful closed-source models such as GPT-4o.
Notably, \method-14B achieves comparable performance with GPT-4o on the OpenHands framework, while on the AutoCoderRover framework, the performance gap widens substantially. 
These observations indicate that while proprietary alignment methods are highly effective within specific agentic workflows, additional training strategies are still essential to enhance cross-framework adaptability.

\begin{table}[t]
\centering
\caption{Alignment Generalization Across OpenHands and AutoCoderRover on SWE-Bench-Lite. \method only use OpenHands framework to construct training dataset.}
\label{table:autocoderover}
% \resizebox{0.8\linewidth}{!}{
\begin{tabular}{clc}
\toprule
{Agentic Framework} & {Model} & {Resolve Rate (\%) $\uparrow$} \\
\midrule
\multirow{3}{*}{OpenHands} & Qwen-14B & 3.7 \\
& \method-14B & 17.7 \\
& GPT-4o & 21.0 \\
\midrule
\multirow{3}{*}{AutoCodeRover} & Qwen-14B & 2.0 \\
& \method-14B & 5.7 \\
& GPT-4o & 30.7 \\
\bottomrule
\end{tabular}
% }
\end{table}

\subsection{Overhead for Execution Validation}
\label{sec:timeefficiency}
Although \method requires minimal training overhead and a small number of data samples, one of the main bottlenecks lies in the inefficiency and significant computational overhead associated with \textbf{execution validation}.
This challenge impacts both the scalability of \method and the feasibility of applying similar alignment training methods to large-scale datasets for software engineering agentic tasks.
During our experiments, substantial computational and storage resources are required. \ding{182} Approximately 1TB of free disk space is needed for intermediate outputs. \ding{183} Using a 16-CPU server, it takes nearly 5 minutes to complete execution validation for a single sample with the official docker platform \cite{swebench, swegym}.

As a result, constructing training datasets and running experimental evaluations involved significant time and resource costs. Scaling alignment methods like \method to larger datasets or more environments will therefore require solving the challenge of execution validation efficiency. Optimizing this step, perhaps by employing lightweight performance estimations or other efficiency strategies, can be critical for future improvements in scalability and practicality.

\subsection{Comparison with Advanced RL Techniques}
\label{sec:onlinerl}

While our study demonstrates the effectiveness of \method, we have not extensively compared it with advanced reinforcement learning (RL) alignment techniques, such as {DeepSeek-R1} \cite{guo2025deepseekr1}. RL-based alignment models, including those trained with methods like GRPO \cite{shao2024deepseekmath}, have shown strong performance in reasoning scenarios such as mathematics and competition-level code tasks. These techniques typically utilize online training environments, leveraging high-quality datasets and carefully designed reward systems. However, this comes at the cost of requiring substantial training resources and computational overhead.  

In contrast, offline alignment techniques \cite{rafailov2024direct}, which is adopted by \method, approximate similar optimization objectives while introducing necessary simplifications. For example, \method leverages pre-generated trajectory datasets and critical node optimization methods instead of conducting resource-intensive online sampling. This design choice reduces computational costs and enables straightforward implementation in constrained environments.

To further explore the potential of advanced RL-based techniques, we perform a simplified experiment using an online iterative DPO approach \cite{pang2024iterative}. We use the additional 50\% on-policy dataset based on \method and results are shown in Table \ref{table:data_scaling_results}. However, results suggest that this method shows limited improvement. A detailed comparison of how \method performs relative to advanced RL alignment techniques in both effectiveness and efficiency remains an open question.

\subsection{Threats to Validity}

In this section, we outline potential threats to the validity of our study and address measures taken to mitigate them.

\paragraph{Threats to Internal Validity}  
Threats to internal validity are related to the dataset collection process and hyperparameter settings used during training. In our experiments, the quality and diversity of the trajectories could influence the model’s ability to generalize across different tasks. Additionally, hyperparameter choices such as the learning rate may impact the final performance. While we conduct ablation studies to ensure robustness, there remains a possibility that certain experimental configurations may bias results.

\paragraph{Threats to External Validity}  
External validity threats are related to the tasks and datasets selected for evaluation. Our study relies on benchmarks such as SWE-Bench-Lite, SWE-Bench-Verified, and HumanEvalFix, which simulate real-world software engineering scenarios. However, these benchmarks may not fully capture the diversity and complexity of all possible software engineering tasks. The results may therefore not generalize to other tasks or repositories beyond those included in the benchmarks.

\paragraph{Threats to Construct Validity}  
Construct validity threats are related to the evaluation metrics used to assess model performance. In this work, we primarily rely on resolve rate, which measures the percentage of successfully completed tasks.  While this metric provides a direct assessment of task success, it may not fully capture other aspects of model performance, such as efficiency, code quality, or usability in real-world scenarios.

\section{Conclusion}

In this paper, we introduce \method, a novel alignment framework designed to address the gap between existing code model training methods and the requirements of real-world software engineering agents. 
\method leverages unique characteristics of software development processes, including the identification of critical decision points, fine-grained alignment and preference optimization, enabling models to better handle complex software engineering tasks within agentic frameworks.
Our experimental results demonstrate that \method achieves state-of-the-art performance across multiple benchmarks designed to simulate real-world software engineering challenges. Additionally, we apply \method to real-world software engineering scenarios, successfully automating the creation of several small applications. 
We are convinced that ongoing advancements in alignment techniques, scalability, and broader applicability will significantly accelerate the realization of fully automated software engineering powered by LLMs.

\bibliographystyle{ACM-Reference-Format}
\bibliography{sample-base}

%%
%% If your work has an appendix, this is the place to put it.

\end{document}